# High Performance Polarization Management Devices Based on Thin-Film Lithium Niobate

Zhongjin Lin[1, 2, #], Yanmei Lin[1, #], Hao Li[1, #], Mengyue Xu[1], Mingbo He[1], Wei Ke[1], Zhaohui Li[1], Dawei Wang[1], X.Steve Yao[3], Siyuan Yu[1], Xinlun Cai[1, *]

1 State Key Laboratory of Optoelectronic Materials and Technologies, School of Electronics and Information Technology, Sun Yat-sen University, Guangzhou 510275, China

2 Department of Electrical and Computer Engineering, The University of British Columbia, Vancouver, BC V6T 1Z4, Canada

3 Photonics Information Innovation Center and Hebei Provincial Center for Optical Sensing Innovations, College of Physics Science and Technology, Hebei University, Baoding 071002, China

# These authors contributed equally

* Corresponding author: caixlun5@mail.sysu.edu.cn

**Abstract:** High-speed polarization management is highly desirable for many applications, such as remote sensing, telecommunication, and medical diagnosis. However, most of the approaches for polarization management rely on bulky optical components that are slow to respond, cumbersome to use, and sometimes with high drive voltages. Here, we overcome these limitations by harnessing photonic integrated circuits based on thin-film lithium niobate platform. We successfully realize a portfolio of thin-film lithium niobate devices for essential polarization management functionalities, including arbitrary polarization generation, fast polarization measurement, polarization scrambling, and automatic polarization control. The present devices feature ultra-fast control speed, low drive voltages, low optical losses and compact footprints. Using these devices, we achieve high fidelity polarization generation with a polarization extinction ratio up to 41.9 dB, fast polarization scrambling with a scrambling rate up to 65 Mrad/s, and endless polarization control with a tracking speed up to 10 Krad/s, all of which are best results in integrated optics. The demonstrated devices unlock a drastically new level of performance and scales in polarization management devices, leading to a paradigm shift in polarization management.

## 1. Introduction

State of polarization (SOP), the vectorial signature of light, is of paramount importance for both fundamental research and practical applications[1, 2]. Unlike insects and some vertebrates that possess sensory mechanisms to perceive polarized light patterns for navigation[3-5], humans rely on specific devices to control and use the SOP of light for widespread applications, including remote sensing[6], telecommunication[7, 8], medical diagnosis[9], and material analysis[10-13]. Polarization management devices capable of fast and dynamic control over the SOP are highly desirable. For example, by analyzing the SOP of optical signal, light remote detection and ranging (LiDAR) systems can reveal the profiles and types of aerosol particles, which is important for monitoring the air pollution and predicting the climate change[14]. In this case, devices for SOP generation and measurement, with high-speed and high-accuracy, are crucial for improving the throughput and spatial-temporal resolutions of SOP LiDAR systems. In optical fiber communication systems, an automatic polarization control device can be used to track and stabilize the fluctuation of SOPs at the receiver end[15, 16]. This approach has the potential to simplify digital signal processing algorithms and reduce the power consumption. In this scenario, the high-speed devices with an unlimited transformation range, or "endless" operation, are required to avoid any interruptions or reset processes.

To date, most of the polarization management devices are based on mechanically rotated wave-plates or fiber-optic coils[17, 18], but these solutions are slow and may introduce instabilities from mechanical vibrations. Higher speed can be achieved with liquid crystal devices[19-21], combining fibers with piezo-electric actuators[22], or combining waveplates with magneto-optic crystals[23]. However, all these approaches rely on bulk-optic components and the control speed is limited to the milli- or micro-second levels. Electro-optic (EO) polarization management in lithium niobate (LN) can obtain very fast control speeds on the order of nano-seconds, where the electrical field applied to the LN waveguide enables modification of SOP through the Pockels effect[15, 16]. While LN devices are attractive for fast polarization management, the performance of these devices is already reaching the physical limits that the conventional LN waveguides can ever support. The current off-the-shelf LN polarization management device is still bulky in size (> 5 cm) and suffer from high drive voltage range (> 100 V)[16, 24], which severely limit their applications in communications and sensing.

Recently, LN-on-insulator (LNOI) has emerged as a promising platform for future EO integrated devices[25-29]. This platform seamlessly combines the superior EO modulation property of LN material with high-index-contrast waveguide structure. As a result, the LNOI devices exhibit much lower drive voltages and much smaller size, compared to their conventional counterparts. Here, we report a portfolio of LNOI-based photonic integrated circuits (PICs) capable of realizing the essential polarization management functionalities, including the arbitrary polarization generation, fast polarization scrambling, fast polarization measurement, and endless automatic polarization control. These devices feature ultra-fast control speed, low drive voltages, low optical losses and compact footprints, many of which exhibit performance well beyond the state-of-the-art.

## Results
### Basic Building Blocks

All of the demonstrated LNOI PICs are based on two fundamental building blocks. The first one is the polarization splitter and rotator (PSR), which maps the orthogonal linearly polarized states in free space, |V⟩ and |H⟩, into two guided modes in two LNOI waveguides, |0⟩ and |1⟩, and vice versa. |0⟩ and |1⟩ represent fundamental TE modes of upper and lower waveguides in Fig.1a, respectively. The second one is a 2×2 Mach-Zehnder interferometer (MZI), composed of two voltage-controlled EO phase shifters and two 3-dB multi-mode interferometer (MMI) couplers (Fig 1b). The relative amplitude and phase between |0⟩ and |1⟩ can be reconfigured by the first and second EO phase shifter. In general, this architecture allows arbitrary unitary transformation to be performed within the PICs, which in turn manipulate the SOP in the output port.

We designed and fabricated the two building blocks on an X-cut LNOI platform, both of which exhibit high performance. The device design and the fabrication process are detailed in Supplementary Material I-IV and XI. The principle of the PSR is based on a mode evolution scheme, which consists of a mode evolution taper, an asymmetric directional coupler, and an edge coupler for off-chip coupling. The PSR features a polarization cross talk of ~ 23 dB, an operation bandwidth of > 70 nm, and an on-chip insertion loss of < 0.12 dB (see Fig. 1c). Moreover, the device also exhibits an off-chip coupling loss of < 1.7 dB for both polarizations with a polarization dependent loss (PDL) of 0.15 dB. The MZI operates in a single-drive push–pull configuration, so that the electric fields induce phase shifts (see Fig. 1d) with an equal magnitude but opposite sign in the two arms. A critical figure of merit for MZI is the half-wave voltage ($V_\pi$), defined as the voltage required to switch the transmission from cross to bar states. For the MZI used in the present devices, the arm length is 1.2 cm. We measure a $V_\pi$ of 2.4 V (see Fig. 1e) and a switching speed of < 5 ns (see Fig. 1f), which facilitates high-speed and low-power operation. Importantly, the MZI also features a low insertion loss of less than 0.4 dB and an extinction ratio (ER) of about ~22.6 dB.

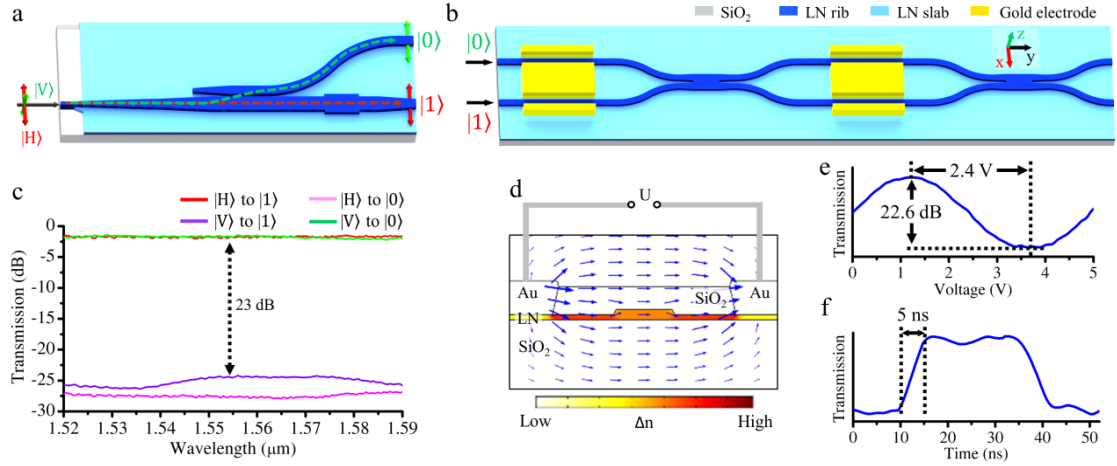

**Fig. 1 Architecture and performance of the fundamental building blocks. a** The schematics of the polarization splitter and rotator (PSR). |V⟩ and |H⟩ represent the vertical and horizontal linearly polarized states in free space, respectively. |0⟩ and |1⟩ represent fundamental TE modes of upper and lower waveguides, respectively. **b** The architecture of 2×2 Mach-Zehnder interferometer (MZI), composed of two electro-optic (EO) phase shifters. **c** The measured transmissions of the fabricated PSR (including the off-chip coupling loss) when injecting horizontal polarization |H⟩ and vertical polarization |V⟩, respectively. **d** The numerical result of the refractive index variation distribution of the lithium niobate (LN) waveguide when applying the voltage to the electrode. **e** The normalized transmission of a MZI as a function of an applied voltage, showing a half-wave voltage of 2.4 V and an extinction ratio of 22.6 dB. **f** The normalized transmission of a MZI as a function of time when applying rectangle wave signals, indicating a fast switching speed of <5 ns.

## Arbitrary SOP generation

Arbitrary SOP generation can be implemented using two PSRs and one MZI, as illustrated in Fig. 2a. We set θ and φ as the phase difference induced by the first and second EO phase shifters, respectively. When the input SOP is set at |H⟩, the normalized Stokes vector **S** of the output SOP can be expressed by

$$\mathbf{S} = (s_0 \quad s_1 \quad s_2 \quad s_3)^T = (1 \quad \cos\theta \quad \sin\theta\cos\varphi \quad \sin\theta\sin\varphi)^T, \quad (1)$$

where $s_0$, $s_1$, $s_2$ and $s_3$ are the four Stokes parameters. The north and south poles of the Poincaré sphere are represented by the points of $s_1=1$ and $s_1=-1$, respectively. The derivation of Equation (1) is in Supplementary Material V. Equation (1) indicates that we can independently control the longitude or latitude position of the output SOP on the Poincaré sphere by θ or φ (inset of Fig. 2a).

We test our arbitrary SOP generating device by applying two triangle wave signals with frequencies of 200 Hz and 10 kHz to θ and φ phase shifters, respectively. The peak-to-peak drive voltages ($V_{pp}$) are set to be 4.8 V, corresponding to $2\pi$ phase shift. In this case, the device is supposed to sample over the entire Poincaré sphere. We use a commercial polarization analyzer (General Photonics PSY 201) to characterize the generated SOP from our device. The waveform of the modulated stokes parameters and the sampling points on the Poincaré sphere are depicted in Fig. 2b and 2c, respectively. In this work, we have addressed the issue of the direct-current (DC) drift in LN device.[30, 31] Therefore, the proposed device can achieve a solid angle repeatability of 0.107 ° in Poincaré sphere. The repeatability of generated SOPs at given voltages is detailed in Supplementary Material V. The polarization extinction ratio (PER) was measured to be 22.8 dB. As expected, the PER of the device coincides with the ER of MZI. We note that the limited PER leads to two un-sampled areas in the vicinity of the north and south poles in Fig. 2c. In principle, the higher PER can be achieved by improving the ER of MZI. However, the ER of MZI is difficult to improve in practice because the fabrication imperfections that can cause the splitting ratio of MMI to deviate from 50:50 always exist.

To fully overcome the restriction of limited ER in MZI, we further demonstrate a multi-stage device, in which additional interferometers function[32] as beam splitters with variable splitting ratio (Fig. 2d). The Poincaré sphere can be fully covered by the SOP generated from this device (Fig. 2e). The measured PER is 41.9 dB, which is the best value in integrated optics[33]. The details of this multi-stage device can be found in Supplementary Material V.

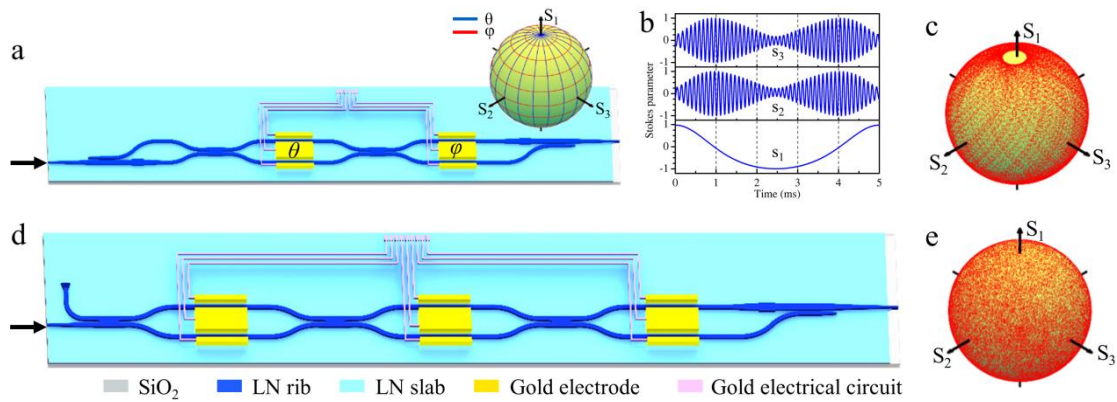

**Fig. 2 Performance of the device for arbitrary SOP generation. a** The schematics of the device for the arbitrary SOP generation. Inset: the position of SOP on the Poincaré sphere when changing θ (blue line) or φ (red line) of the electro-optic (EO) phase shifters. **b** The measured Stokes vectors as a function of time when applying two triangle wave signals with frequencies of 200 Hz and 10 kHz to θ and φ EO phase shifters, respectively. **c** The measured Stokes vectors on the Poincaré sphere of the output from the device, indicating a polarization extinction ratio (PER) of 22.8 dB. **d** The schematics of a multi-stage device with an improved PER. **e** The measured Stokes vectors of the output from the multi-stage device, indicating a PER of 41.9 dB.

**Polarization scrambling**

We programmed our arbitrary SOP generating device to implement fast polarization scrambling (Fig. 3a), which can be used to mitigate polarization related impairments in optical fiber communication and sensing systems. An important figure of merit for polarization scrambling device is the scrambling rate, defined as the rate of polarization change on the surface of Poincaré sphere. The scrambling rate needs to be fast enough so that the average polarization over certain period of time effectively covers the entire surface of Poincaré sphere. Fig.3a depicts the experimental setup for testing the scrambling rate. We applied two triangle waves with frequencies of 4.242 MHz and 6 MHz to θ and φ phase shifters, respectively. In this case, the corresponding scrambling rate is 65 Mrad/s (see Equation (S13) of Supplementary Material VII). A commercial polarization analyzer was used to measure the degree of polarization (DOP) of the output from our device. Fig. 3b presents the DOP as a function of the detector integration time at a wavelength of 1550 nm. The result indicated that a DOP of 1 % can be achieved at a detector integration time of only 1 μs. The DOP as a function of the scrambling rate is provided in Supplementary Material VII. Fig. 3c shows the measured DOP at different wavelength for a detector integration time of 1 μs, indicating broadband operation of the device over 70 nm. These results demonstrate that our scrambling device significantly outperforms the best commercial scrambling device[34] with a maximum scrambling rate of 50 Mrad/s.

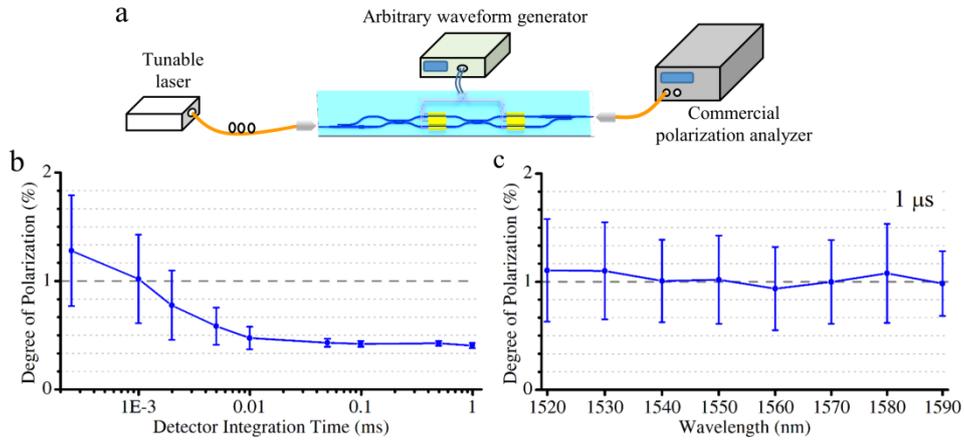

**Fig. 3 Performance of the polarization scrambling device. a** The experiment setup for characterizing the performance of the polarization scrambling device. **b** The degree of polarization (DOP) as a function of the detector integration time at a wavelength of 1550 nm. **c** The DOP as a function of the wavelength when the detector integration time is equal to 1 μs.

**Fast polarization measurement**

We realized fast and accurate SOP measurement using a LNOI PIC composed of a PSR, a MZI and a flip-chip bonded photodetector (PD), as depicted in Fig. 4a. It has an architecture similar to the SOP generating device, but with the PD for monitoring the optical intensity. In general, the SOP can be measured by making four projective intensity measurements onto four sets of predetermined measurement basis[35-37] which are parameterized by a 4×4 measurement matrix, **W**. The Stocks vector is then given by **S**=**W**$^{-1}$**I**, where **I**= ($I_1$, $I_2$, $I_3$, $I_4$) is a 4-dimensional vector representing the results of projective intensity measurements. In our case, $I_1$, $I_2$, $I_3$ and $I_4$ are measured by the PD and the matrix elements of **W** are determined by setting different values on θ and φ.

We adopted the "optimal frames" protocol, which has been developed as an effective method to minimize the influence of the noise (see Supplementary Material VI)[35], to perform fast SOP measurement (Fig. 4a). An input optical signal with rapidly and randomly varying SOP, generated from a commercial polarization synthesizer/analyzer, was used to interrogate the performance of our device. Fig. 4b shows the comparison of the measured SOP from our device and the original input SOP, both of which are parameterized by Stokes vector elements. The results from our device exhibited high level of accuracy with the RMS deviation of 1.58°. Moreover, the sampling rate of the device can go up to 250 KS/s without compromising the detection accuracy, thanks to the ultra-fast response of the LNOI material. Our device can also be used to measure the DOP of the input light beam, which is detailed in Supplementary Material VI.

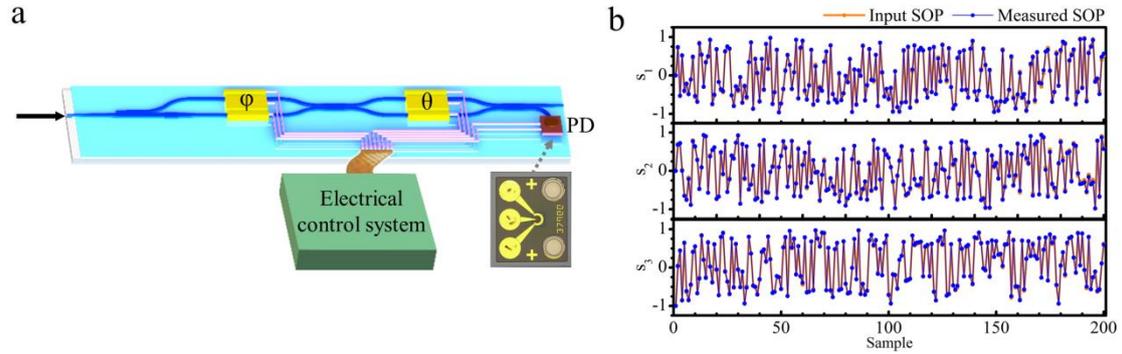

**Fig. 4 Performance of the device for fast polarization measurement. a** The experiment setup used for characterizing the fast polarization measuring device. Inset: the microscope image of a flip-chip bonded photodetector (PD). According to the optical intensities recorded by the PD, the input state of polarization (SOP) can be reconstructed. **b** The measured results (blue line) from the device and the corresponding input polarization states (orange line).

**Endless automatic polarization control**

An essential requirement of automatic polarization control is that it must be able to operate in an endless way. In other words, the polarization control device is capable of continuously tracking the rapidly varying changes of all possible SOPs, even if the SOPs wander infinite times around the Poincaré sphere[16]. Although the PIC with one PSR and one MZI which includes two EO phase shifters is sufficient for transforming any input SOP to the TE mode[38], the reset process is inevitable when the drive voltages reach the boundaries (see Supplementary Material VIII), resulting in momentary SOP mismatch and data interruption. To achieve polarization control with endless and seamless operation, multi-stage MZIs with at least four EO phase shifters are needed.

We developed the LNOI endless automatic polarization control device, composed of a PSR, a flip-chip bonded PD, and a multi-stage MZI with four EO phase shifters, as depicted in Fig. 5a. The PD was used to monitor the optical power from one of the output waveguide (port 2 in Fig. 5a), which was further used as a feedback to control the voltages applied onto the four EO phase shifters. A gradient algorithm was digitally implemented in a field-programmable gate array (FPGA) for fast execution. The algorithm was design to minimize the power received by the PD under all possible input SOPs. Thus all of the power of the input signal with arbitrary SOP would be transferred to the other output waveguide with TE polarization (port 1 in Fig. 5a). Fig. 5b depicted the micrograph of the fabricated devices. The device is folded to reduce the total length, and the footprint is about 1.5 cm × 0.3 cm. We measured a low on-chip insertion loss of 0.92 dB for the device.

We use a commercial polarization scrambling device to test the performance of our polarization control device. To verify the endless operation capability of our device, the input SOPs were first programmed to endlessly rotate around three different orthodromes on the Poincaré sphere at the speed of 2 rad/s (Fig. 5c insets). We perform tracking experiment for more than 1 hour for each great orthodrome, corresponding to a SOP change of more than 7000 rad. Fig. 5c shows that the normalized optical power received by the PD was always kept at a low level of

< 0.1%, indicating the successful endless tracking operation. To further test the tracking rate of our device, we reprogrammed the commercial polarization scrambler to generate rapidly and randomly varying SOPs at different scrambling rate, and then we use our device to track and stabilize the scrambled SOPs. As shown in Fig. 5d, the normalized power received by the flip-chip bonded PD could still be maintain at around 1%, even at the SOP changing rate of 10 Krad/s. This is by far the highest endless tracking rate reported in integrated photonics. It should be noted that the faster tracking rate of 1 Mrad/s can be achieved by simply upgrading the electronic control system.

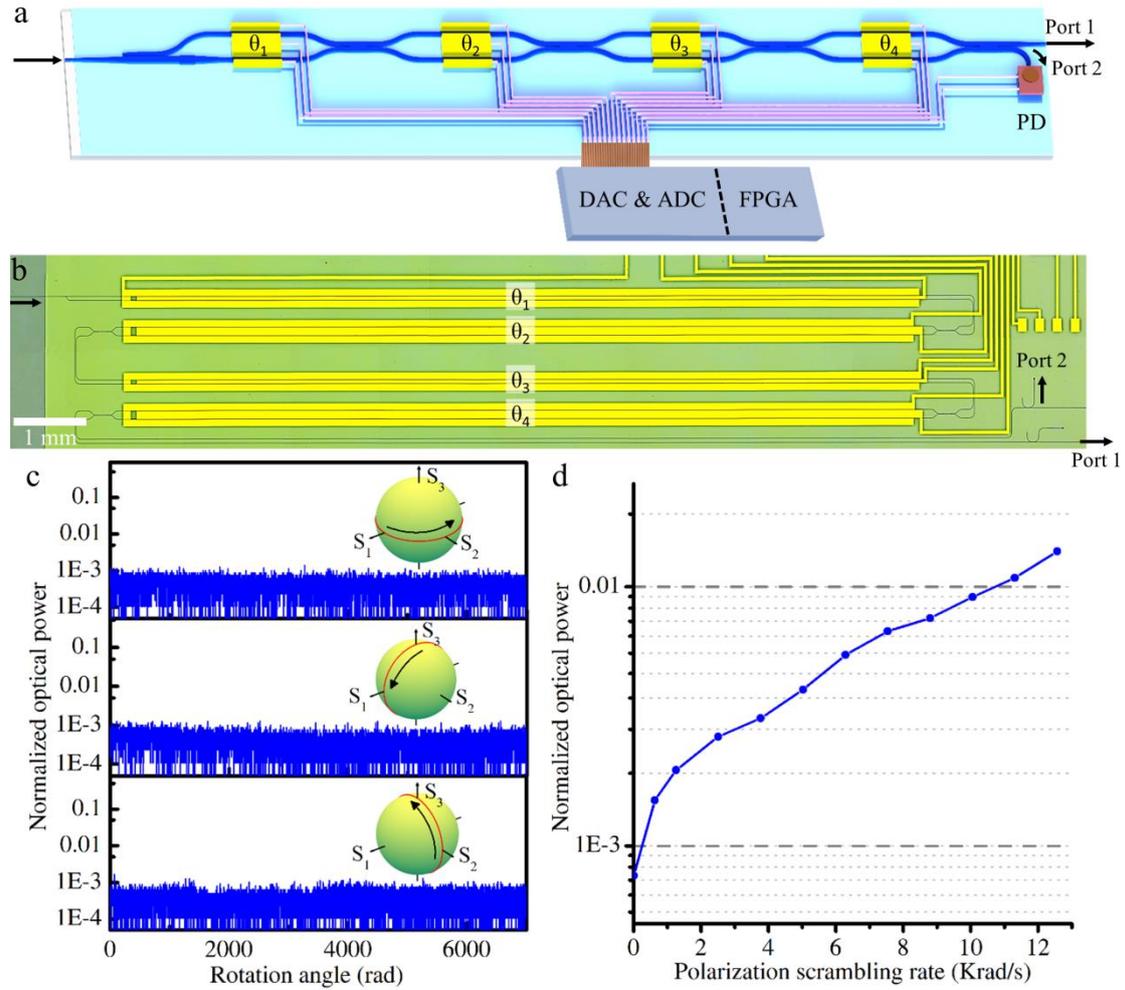

**Fig. 5 The performance of the endless automatic polarization controlling device. a** The schematics of the lithium niobate-on-insulator (LNOI) endless automatic polarization controlling device. DAC: digital-to-analog converter. ADC: analog-to-digital converter. FPGA: field-programmable gate array. PD: flip-chip bonded photodetector. **b** The microscope image of the fabricated device. $\theta_1$, $\theta_2$, $\theta_3$ and $\theta_4$ in (a) and (b) mark the phase shifters. **c** The normalized optical power of the PD as a function of the rotation angle of the input polarization. Insets: the rotation trajectories of the input polarization states for demonstrating the capability of the endless operation. **d** The normalized optical power of the feedback signal as a function of the polarization scrambling rate.

## Discussion

We have successfully demonstrated high-performance LNOI-based PICs promising for polarization management applications. The LNOI material provides a stable, compact and robust platform to implement high-speed and exquisite polarization management. Sophisticated devices can be fabricated in a mass-produced way by standard semiconductor process, such as lithography, etching and metal patterning. Indeed, this approach brings new levels of performance, functionality and scalability to polarization management.

It should be noted that polarization management functionalities have also been implemented on other material platform, such as silicon-on-insulator (SOI)[33, 39-43], indium phosphide (InP)[44, 45], plasmonics[46] and conventional LN[24]. In Table 1, we compare the performance metrics of our device to the state-of-the-art. Clearly, the present device features the fastest response speed and lowest drive voltage. To the best of our knowledge, these are also the records for all polarization devices. In particular, the low drive voltage of our device is highly attractive for high-speed and power-efficient operation. The low optical loss of our device is also very appealing. For example, the endless polarization control device features an on-chip insertion loss of only 0.92 dB. Therefore, it can be monolithically integrated with other LNOI devices and balance PDs to form integrated coherent receivers capable of polarization de-multiplexing, bringing new possibilities to future high bandwidth and low power consumption optical networks. Importantly, the LNOI platform provides pure phase modulation which means the intensity of the light does not change with an external modulation voltage. This fundamentally avoids the activation loss commonly observed in silicon and InP devices. In fact, the measured activation loss in our device is negligible (see Supplementary Material V). Furthermore, LNOI platform support a wide transparent window from 400 nm to 5000 nm[47], compared to other material systems. This also opens up new application opportunities in areas like biology, chemistry, medicine, remote-sensing, and astronomy.

**Table 1** The comparison of several performance metrics of the active integrated polarization devices.

| Platform | Principle | $V_\pi$ (V) | Tuning Power (W) | Response time (ns) | Optical losses[a] (dB) | Transparent window (μm) |
|---|---|---|---|---|---|---|
| SOI[38, 39, 48] | Thermo-optics effect | < 10 | < 1 | > 5 ×10$^4$ | NA | 1-5[49] |
| SOI[8] | Plasma dispersion effect | 7.07 | NA | < 2.5 | 5 | 1-5[49] |
| InP[44] | Plasma dispersion effect | < 3 | NA | NA | 5.5 | 1.1-1.6[50] |
| Plasmonic[46] | Faraday effect | NA | NA | < 1000 | 2.5[b] | NA |
| Ti:LiNbO$_3$[24,c] | Pockels effect | ~28 | ~5 | < 10 | NA | 0.4-5[47] |
| This work | Pockels effect | ~2.4 | < 10$^{-3}$ | < 5 | 0.52[d] 0.92[e] | 0.4-5[47] |

NA: not available.

a: On-chip insertion loss.

b: Only generating an arbitrary linear polarization state.

c: Including rotatable waveplates.

d: Including a PSR, and a single-stage MZI with two EO phase shifters

e: Including a PSR, and a multi-stage MZI with four EO phase shifters

**Acknowledgements**

The device fabrication is performed at the State Key Laboratory of Optoelectronic Materials and Technologies, School of Electronics and Information Technology, Sun Yat-sen University. We thank Lidan Zhou, Shenqiang Gao and Lin Liu for their technical support. This work was supported by the National Key Research and Development

**Author contributions**

Z. L. and X. C. prepared the manuscript in discussion with all authors. Z. L designed the devices, and the control system including the algorithm. Z. L and Y. L. performed the experiments and analyzed the data. H. L. fabricated the devices with the help from M. H and W. K. Z. L, D. W, X. Y and S. Y provided some suggestion about the experiment and revised the manuscript. X.C supervised the project.

**Data availability**

The data that support the plots within this paper and other findings of this study are available from the corresponding author upon reasonable request.

**Conflict of interest**

The authors declare that they have no conflict of interest.

**Supplementary information** is available for this paper at https://doi.org/.

# Supporting information for "High Performance Polarization Management Devices Based on Thin-Film Lithium Niobate"

Zhongjin Lin[1,2,#], Yanmei Lin[1,#], Hao Li[1,#], Mengyue Xu[1], Mingbo He[1], Wei Ke[1], Zhaohui Li[1], Dawei Wang[1], X.Steve Yao[3], Siyuan Yu[1], Xinlun Cai[1,*]

1 State Key Laboratory of Optoelectronic Materials and Technologies, School of Electronics and Information Technology, Sun Yat-sen University, Guangzhou 510275, China

2 Department of Electrical and Computer Engineering, The University of British Columbia, Vancouver, BC V6T 1Z4, Canada

3 Photonics Information Innovation Center and Hebei Provincial Center for Optical Sensing Innovations, College of Physics Science and Technology, Hebei University, Baoding 071002, China

# These authors contributed equally

* Corresponding author: caixlun5@mail.sysu.edu.cn

## Content

## I. The performance of the edge coupler

The schematic of the proposed polarization-independent edge coupler is presented in Fig. S1a. Both rib and slab portions of the waveguide utilize the architecture of gradual tapering. The top (rib) taper evolves from a width of 800 nm down to 80 nm with a height $h_{rib}$ of 180 nm over a 150 μm length. The bottom (slab) layer taper narrows down laterally from 6 μm to a tip of 80 nm wide with a slab height $h_{slab}$ of 180 nm. At the end, a strip waveguide with a width of 80 nm is used to match the optical mode of fiber.

The microscope and scanning electron images of the fabricated edge coupler are shown in the inset of Fig. S1b. At the wavelength of 1550 nm, the off-chip coupling losses of the edge coupler are 1.7 dB and 1.55 dB when injecting horizontal polarization and vertical polarization, respectively (Fig. S1), indicating a low polarization-dependent loss of 0.15 dB. The 3-dB bandwidth covers the range of 70 nm.

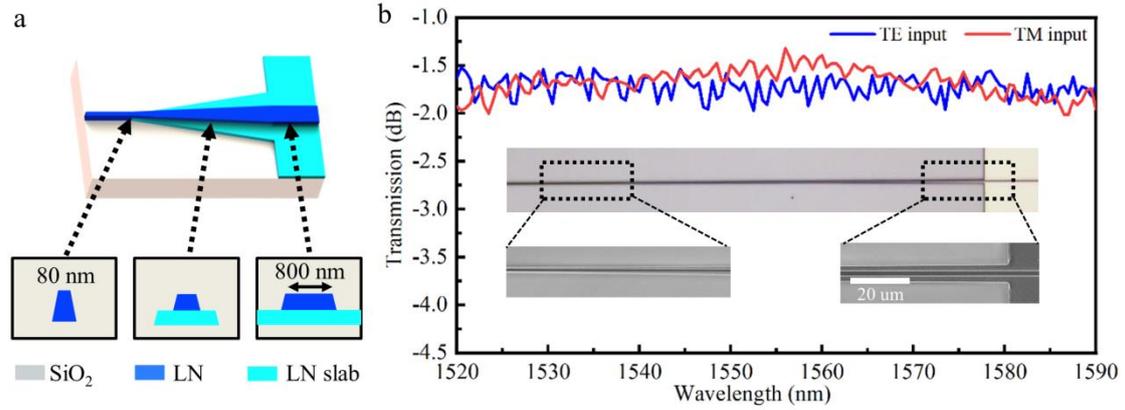

Fig. S1 **The performance of the polarization-independent edge coupler.** (a) The schematic of the proposed edge coupler. Insets: the profiles of different sections. (b) The measured transmissions of the fabricated edge coupler when injecting horizontal polarization (TE, blue line) and vertical polarization (TM, red line), respectively. Top of inset: the microscope image of the fabricated edge coupler. Bottom of inset: the scanning electron image of the special section of the fabricated edge coupler.

## II. The design of the polarization splitter and rotator

The schematic of the polarization splitter and rotator (PSR) is presented in Fig. S2a. It includes an adiabatic taper (AT), an asymmetrical directional coupler (ADC), and a multimode interferometer mode filter (MMIF). The AT is split into three sections. The widths of the three sections are 0.8 μm, 1.1 μm, 1.7 μm, and 2.16 μm, respectively. The corresponding lengths are 50 μm, 250 μm, and 50 μm, respectively. The length of the coupled region of the ADC is 34 μm. According to the phase-matching condition, the widths for the straight bus and access waveguides of the ADC are 2.16 μm and 0.8 μm, respectively. The width and length of the MMIF are 6 μm and 53.5 μm, respectively. Fig. S2b presents the microscope image of the fabricated PSR. We will introduce the principle of the PSR in the following.

The insets of Fig. S2b provide the mode evolution at different sections of the proposed device. When an arbitrary polarization state is injected into the AT from the right side, the $TM_0$ mode is converted into the $TE_1$ mode, but the launched $TE_0$ mode does not undergo mode conversion. Then, the ADC is designed to split the $TE_0$ and $TE_1$ modes into the straight bus and access waveguides, respectively, and simultaneously converts the $TE_1$ mode to the $TE_0$ mode. The MMIF, connecting with the straight bus waveguide, is used for filtering out the residual $TE_1$ mode so that a high extinction ratio can be achieved when the $TM_0$ mode is launched. All the structures, including the AT section which is used for converting the TM mode to TE mode, are fabricated in a single lithography step, which indicates a high fabrication tolerance.

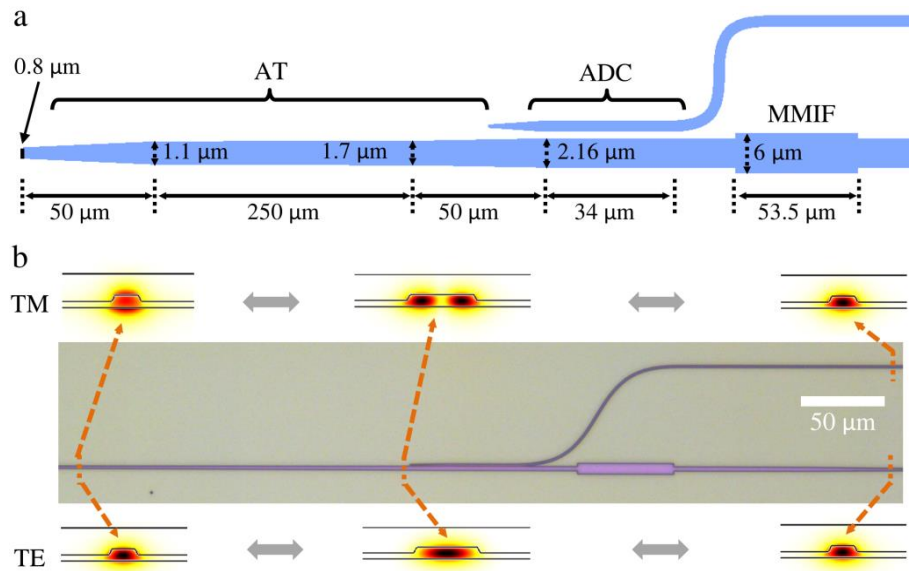

Fig. S2 **The schematic of the polarization splitter and rotator.** (a) The design of the polarization splitter and rotator. (b) The microscope image of the fabricated polarization splitter and rotator. Top: the evolution from TE to TE modes. Bottom: the evolution from TM to TE modes.

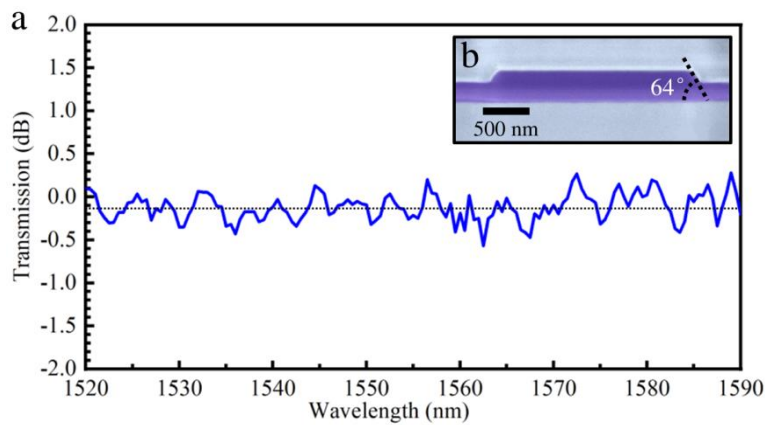

Fig. S3 **The propagation loss of the waveguide.** (a) The transmission of the 2 μm-width waveguides with a length of 1 cm. (b) The scanning electron image of the profile of the fabricated optical waveguide, showing a sidewall angle of 64 °.

## III. The propagation losses of waveguide

Figure S3a shows that the propagation loss of the 2 μm-width waveguides is near 0.11 dB/cm. The scanning electron image of the profile of the optical waveguide presented in Fig. S3b indicates that the sidewall angle of the optical waveguide is equal to 64 °.

## IV. The optimization of the gold electrodes

The profile of the electrodes is shown in Fig. S4a. The optical rid waveguides are located in the dielectric gaps between gold electrodes. The top width of the rid waveguide is 2 μm. The distance between the two electrodes is set at $D_m$. For the voltage-length product ($V_\pi L$), $D_m$ should be as small as possible (red line of Fig. S4b). While a small $D_m$ would result in a high additional optical Ohmic loss (blue line of Fig. S4b). The $D_m$ set at 6 μm can well balance the trade-off.

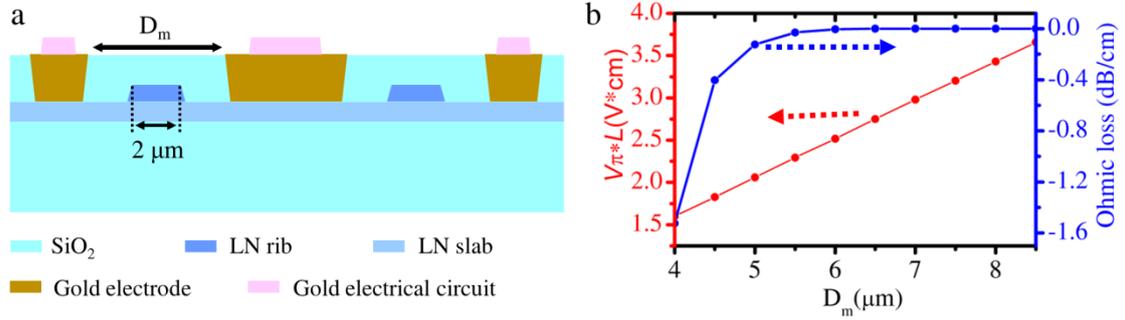

Fig. S4 **The design of the gold electrodes for the phase shifters.** (a) The profile of the phase shifters. (b) The simulation results of the voltage-length product and the Ohmic loss as the functions of the gap ($D_m$) between two electrodes.

## V. The principle and experiment of the arbitrary polarization generation

### Principle

Arbitrary SoP generation can be implemented using two PSRs, a Mach-Zehnder interferometer (MZI), as illustrated in Fig. S5a. Setting the input optical electrical field at $|1\rangle$, the two components $|H\rangle$ and $|V\rangle$ of the output light can be given by,

$$\begin{pmatrix} |V\rangle \\ |H\rangle \end{pmatrix} = C_\varphi \times C_M \times C_\theta \times C_M \times \begin{pmatrix} |1\rangle \\ 0 \end{pmatrix}, \quad (S1)$$

where $C_M$, $C_\theta$ and $C_\varphi$ are the transfer matrices of 3-dB 2×2 multimode interferometer (MMI) coupler, first and second electro-optic (EO) phase shifters, respectively. $C_M$, $C_\theta$ and $C_\varphi$ can be expressed as:

$$C_M = \frac{\sqrt{2}}{2}\begin{pmatrix} 1 & e^{i\frac{\pi}{2}} \\ e^{i\frac{\pi}{2}} & 1 \end{pmatrix}, \quad C_\theta = \begin{pmatrix} e^{i\frac{\theta}{2}} & 0 \\ 0 & e^{-i\frac{\theta}{2}} \end{pmatrix}, \quad C_\varphi = \begin{pmatrix} e^{i\frac{\varphi}{2}} & 0 \\ 0 & e^{-i\frac{\varphi}{2}} \end{pmatrix}. \quad (S2)$$

where $\theta$ and $\varphi$ are the phase shifts of the two EO phase shifters, respectively. The relation between the $|H\rangle$, $|V\rangle$ and the output Stokes vector $(S_0, S_1, S_2, S_3)^T$ can be given by,

$$S_0 = \langle H|H\rangle + \langle V|V\rangle \quad (S3)$$
$$S_1 = \langle H|H\rangle - \langle V|V\rangle, \quad (S4)$$
$$S_2 = \langle V|H\rangle + \langle H|V\rangle, \quad (S5)$$

$$S_3 = i(\langle V|H\rangle - \langle H|V\rangle). \tag{S6}$$

Based on Eqs. S1-S6, we can obtain that,

$$\begin{pmatrix} S_0 & S_1 & S_2 & S_3 \end{pmatrix}^T = \begin{pmatrix} 1 & \cos\theta & \sin\theta\cos\varphi & \sin\theta\sin\varphi \end{pmatrix}^T. \tag{S7}$$

Therefore, controlling the voltages applied to the two EO phase shifters, we can generate an arbitrary pure polarization state. Fig. S5b-S5d presents the microscope images of the fabricated device.

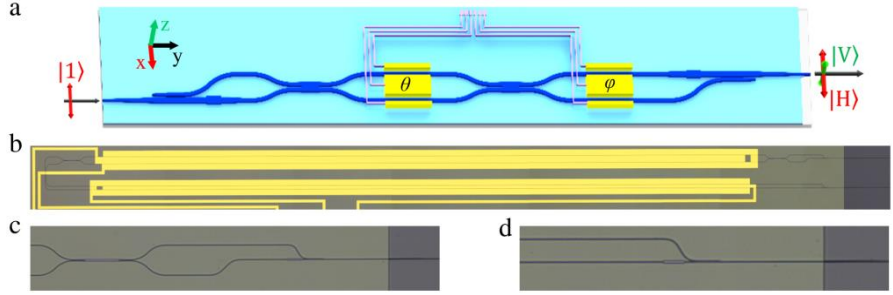

Fig. S5 **The model of the proposed polarization generating device.** (a) The schematic of the single-stage device. (b) The microscope image of the fabricated device. (c) and (d) are the enlarged images of the optical input and output ports, respectively.

In the above, we only consider the case of that the power splitting ratio of MMI coupler is equal to 50:50. However, in practice, the fabrication imperfections always cause the splitting ratio of MMI coupler to deviate from 50:50. To analyze this case, we set the power splitting ratio of MMI coupler at $\tau^2 : \kappa^2$ (where $\tau^2 + \kappa^2 = 1$). The transfer matrix of MMI coupler can be updated as,

$$C_{M'} = \begin{pmatrix} \tau & \kappa e^{i\frac{\pi}{2}} \\ \kappa e^{i\frac{\pi}{2}} & \tau \end{pmatrix} \tag{S8}$$

Based on Eqs. S1, S3-S5 and S8, we can obtain that

$$S_1 = 4\kappa^2\tau^2 \cos\theta - (\tau^2 - \kappa^2)^2. \tag{S9}$$

Therefore, in this case, the maximum polarization extinction ratio (PER) near $S_1=1$ can be given by,

$$\text{PER} = -10\log_{10}\frac{1-\max(S_1)}{1+\max(S_1)} = -10\log_{10}\frac{(\tau^2-\kappa^2)^2}{4\kappa^2\tau^2}. \tag{S10}$$

Equation S10 shows that the PER is finite when $\tau$ is not equal to $\kappa$. To fully overcome this restriction, we further demonstrate a multi-stage device, in which additional interferometers function as beam splitters with variable splitting ratio (Fig. S6). In this case, we can obtain that,

$$S_1 = b^2\{\sin\vartheta \sin\phi + a[\cos\vartheta + \cos\vartheta\cos\phi + \cos\phi]\} - a^3 \tag{S11}$$

where $a = \tau^2 - \kappa^2$, $b = 2\kappa\tau$. $\vartheta$ and $\phi$ are the phase shifts of two phase shifters, respectively. $a[\cos\vartheta + \cos\vartheta\cos\phi + \cos\phi]$ allows the multi-stage device to achieve a higher PER.

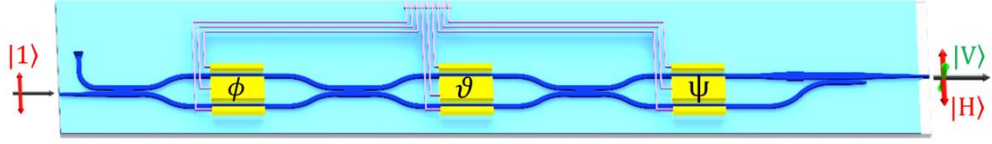

Fig. S6 The schematic of the multi-stage device. $\vartheta$, $\phi$ and $\psi$ are the phase shifts of three phase shifters, respectively.

**Experiment**

Using a commercial polarization synthesizer/analyzer (General Photonics, PSY 201), we can characterize the performance of our device. Here, we would demonstrate that our device can generate highly repeatable polarization states at 0°, ±45°, 90° and LHC and RHC across a Poincaré sphere. Fig. S7a shows the repeatability of the six states on a Poincaré sphere by repeatedly switching among these states 200 times. We can observe that the six states (six red dots) are highly repeatable, with no discernible differences. It is worth noting that because the switching speed is so fast, the switching trajectory among different states cannot be captured by the commercial polarization synthesizer/analyzer. One of the generated SOPs presented in Fig. S7b indicates a high repeatability of the polarization generation. The other five generated SOPs also exhibit the same high repeatability. The solid angle between the two adjacent states is shown in Fig. S7c. The measured solid angle repeatability is 0.107°.

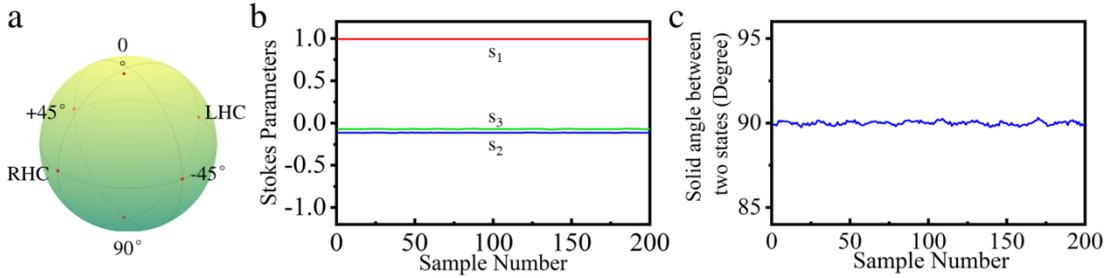

Fig. S7 **Repeatability measurement of our device by switching the generated SOPs among six states.** (a) The six generated SOPs (six red dots) on a Poincaré sphere. (b) Stokes parameters of one of the generated SOPs as a function of sample points. (c) Solid angle between two switching states as a function of sample points.

Activation loss measures the additional insertion loss caused in activating the device. It is defined as the difference of the maximum and minimum insertion losses of the device considering all possible activation conditions. The LNOI platform provides pure phase modulation which means the intensity of the light does not change with an external modulation voltage. This fundamentally avoids the activation loss commonly observed in silicon and InP devices. In fact, the measured activation loss in our device can be negligible (see Fig. S8).

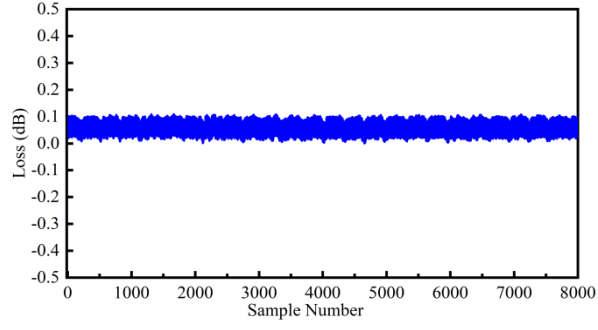

Fig. S8 The loss of our device when applying different voltages.

## VI. The principle and experiment of the polarization measurement

### Principle

The proposed polarization measuring device is composed of a PSR, a MZI and a flip-chip bonded photodetector (PD), as depicted in Fig. S9a. It has an architecture similar to the SoP generating device, but with the PD for monitoring the optical intensity. The optical electrical field ($|E_{PD}\rangle$) received by PD can be given by,

$$|E_{PD}\rangle = (1 \quad 0) \times C_M \times C_\theta \times C_M \times C_\varphi \times \begin{pmatrix} |V\rangle \\ |H\rangle \end{pmatrix}, \quad (S12)$$

where $|V\rangle$ and $|H\rangle$ are two components of the input light. Based on Eqs. S2-S6 and S12, the relation between the optical intensity ($I_{PD}$) and the input Stokes vector ($S'_0$, $S'_1$, $S'_2$, $S'_3$)$^T$ can be given by,

$$I_{PD} = \langle E_{PD}|E_{PD}\rangle = \frac{1}{2}(S'_0 - S'_1\cos\theta + S'_2\sin\theta\cos\varphi + S'_3\sin\theta\sin\varphi). \quad (S13)$$

To reconstruct a full-Stokes vector, the proposed device needs to perform four optical intensity measurements ($I_1$, $I_2$, $I_3$ and $I_4$), which correspond to four drive voltage combinations of $\theta$ and $\varphi$ phase shifters. According to Eq. S13, the input full-Stokes vector can be related to $I_1$, $I_2$, $I_3$ and $I_4$ by,

$$\begin{pmatrix} I_1 \\ I_2 \\ I_3 \\ I_4 \end{pmatrix} = \mathbf{W} \times \begin{pmatrix} S'_0 \\ S'_1 \\ S'_2 \\ S'_3 \end{pmatrix} = \frac{1}{2}\begin{pmatrix} 1 & -\cos\theta_1 & \sin\theta_1\cos\varphi_1 & \sin\theta_1\sin\varphi_1 \\ 1 & -\cos\theta_2 & \sin\theta_2\cos\varphi_2 & \sin\theta_2\sin\varphi_2 \\ 1 & -\cos\theta_3 & \sin\theta_3\cos\varphi_3 & \sin\theta_3\sin\varphi_3 \\ 1 & -\cos\theta_4 & \sin\theta_4\cos\varphi_4 & \sin\theta_4\sin\varphi_4 \end{pmatrix} \times \begin{pmatrix} S'_0 \\ S'_1 \\ S'_2 \\ S'_3 \end{pmatrix}, \quad (S14)$$

where $\theta_i$ and $\varphi_i$ are the phase shifts of two EO phase shifters when recording the i[th] optical power, respectively. If the rank of the analysis matrix $\mathbf{W}$ is equal to 4, the proposed device can be used to reconstruct the full-Stokes vector. The reconstructed accuracy is mainly determined by the noise of the PD and the analysis matrix $\mathbf{W}$. Comparing with the cost of decreasing the noise of the PD, the cost of optimizing the analysis matrix $\mathbf{W}$ can be ignored. For example, we only need to change the drive voltages of two phase shifters to optimize the analysis matrix. According to the previous theoretical analysis[1], the analysis matrix $\mathbf{W}$ is optimal when the analysis matrix $\mathbf{W}$ has the properties of that:

$$\mathbf{W} = \frac{1}{2}\begin{pmatrix} 1 & 1/\sqrt{3} & 1/\sqrt{3} & 1/\sqrt{3} \\ 1 & -1/\sqrt{3} & -1/\sqrt{3} & 1/\sqrt{3} \\ 1 & -1/\sqrt{3} & 1/\sqrt{3} & -1/\sqrt{3} \\ 1 & 1/\sqrt{3} & -1/\sqrt{3} & -1/\sqrt{3} \end{pmatrix}, \quad (S15)$$

or

$$\mathbf{W} = \frac{1}{2}\begin{pmatrix} 1 & 1/\sqrt{3} & 1/\sqrt{3} & 1/\sqrt{3} \\ 1 & -1/\sqrt{3} & -1/\sqrt{3} & 1/\sqrt{3} \\ 1 & -1/\sqrt{3} & 1/\sqrt{3} & -1/\sqrt{3} \\ 1 & 1/\sqrt{3} & -1/\sqrt{3} & -1/\sqrt{3} \end{pmatrix}. \quad (S16)$$

To obtain the properties of Eqs. S15 and S16, $\theta_i$ (where $i$ =1, 2, 3, and 4) can be equal to $\pm\arccos(\frac{1}{\sqrt{3}})$ or $\pm[\pi - \arccos\left(\frac{1}{\sqrt{3}}\right)]$, and $\varphi_i$ can be equal to $\pm\pi/4$, $\pm 3\pi/4$. $\theta_i$ and $\varphi_i$ presented in Fig. S9b are two examples that allow us to obtain the optimal analysis matrix.

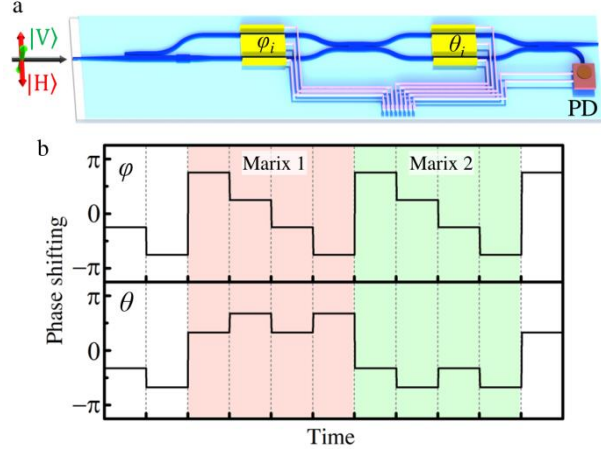

Fig. S9 **The model of the proposed polarization measuring device.** (a) The schematic of the proposed polarization measuring device. (b) The phase shifts of the $\theta$ and $\varphi$ phase shifters when measuring the input polarization states.

**Experiment**

Figure S20 presents the experiment setup for demonstrating our polarization measuring device. The light source consists of a tunable laser and an erbium-doped fiber amplifier (EDFA). Controlling the output optical power ratio between the tunable laser and EDFA, we can change the degree of polarization (DOP) of light source, because the light generated by EDFA is unpolarized and the light generated by the tunable laser is completely polarized. When measuring the results shown in Fig. 4 of the main text, we turned off the EDFA so that the light injected into the photonic chip is completely polarized. The input polarization states shown in Fig. 4 of the main text were randomly generated by a commercial polarization synthesizer/analyzer (General Photonics, PSY 201).

Next, we show that our device can also be used to measure the input light beam whose DOP is lower than 100%. In this case, the commercial polarization synthesizer/analyzer was used to measure the DOP of light which comes from the tunable laser and EDFA. Fig. S10b compares the DOP measured by our device and the commercial one, indicating that RMS deviation of the measured DOP from our device can achieve 2.5%.

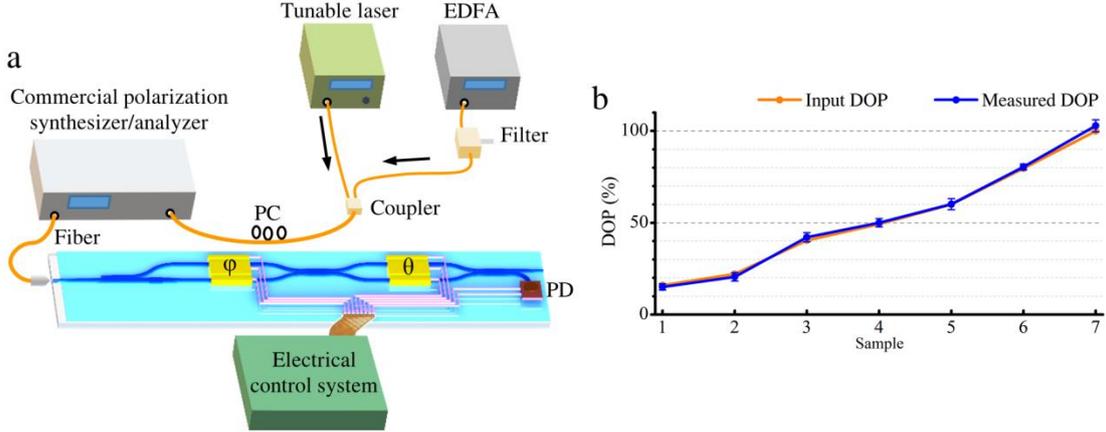

Fig. S10 **The experiment for demonstrating the proposed polarization measuring device.** (a) The schematic of experiment setup for demonstrating the proposed polarization measuring device. EDFA: erbium-doped fiber amplifier. PC: polarization controller. (b) The measured degree of polarization (DOP) from our device and the corresponding input DOP (orange line).

## VII. The experiment for polarization scrambling

The experimental setup for demonstrating polarization scrambling is presented in Fig. S11a. The light is emitted from a tunable laser, and passes through a polarization controller and then is injected into the proposed device by a lens fiber. Our device is controlled by a dual-channel arbitrary waveform generator. The arbitrary waveform generator can provide two triangle wave signals with different frequencies. The frequency ratio of the output signals was set at $1:\sqrt{2}$. The scrambling rate can be controlled by the frequencies of two channels. According to Eq. S7, the scrambling rate can be calculated by:

$$\text{Scrambling rate} = \sqrt{\left(\theta_{range} \times f_\theta\right)^2 + \left(\varphi_{range} \times f_\varphi\right)^2} \qquad \text{(S17)}$$

where $\theta_{range}$ and $\varphi_{range}$ are the phase shifting ranges of the $\theta$ and $\varphi$ phase shifters, respectively. $f_\theta$ and $f_\varphi$ are the frequencies of the triangle wave signals applied to $\theta$ and $\varphi$ phase shifters, respectively. Here, we set $\theta_{range}=4\pi$ and $\varphi_{range}=2\pi$.

A commercial polarization analyzer (General Photonics, PSY 201) was used to characterize the generated polarization state. Fig. S11b provides the degree of polarization of the output light as a function of the scrambling rate of the fabricated polarization scrambling device when the detector integration time is set at 5 μs.

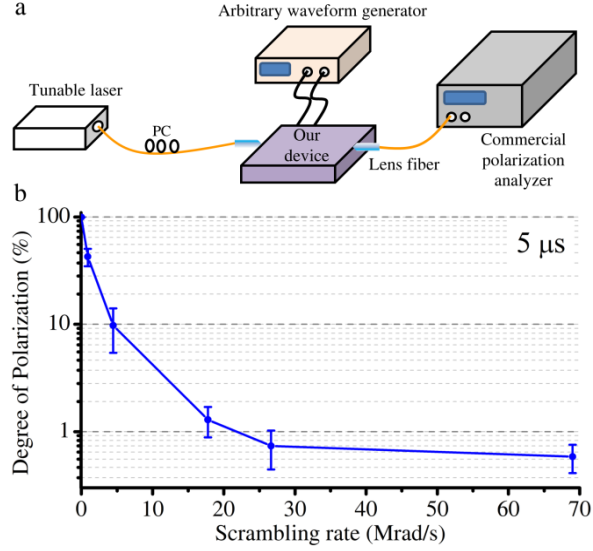

Fig. S11 **The experiment for polarization scrambling.** (a) The schematic of the experimental setup for demonstrating polarization scrambling. PC: fiber polarization controller. (b) The degree of polarization of the output light as a function of the scrambling rate of the fabricated device when the detector integration time is 5 μs.

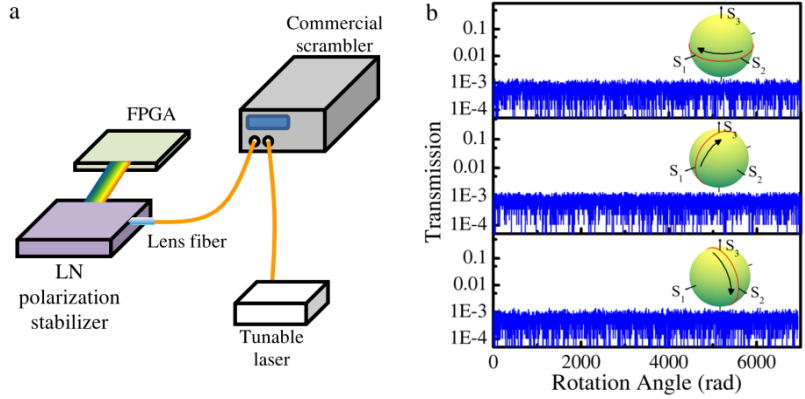

Fig. S12 **The experiment for demonstrating polarization control.** (a) The experiment setup for characterizing our device. (b) The transmission of the feedback signal as a function of the rotation angle of the input polarization. Insets: the trajectories of the input polarization states.

## VIII. The experiment for demonstrating polarization controlling device

The experiment setup of characterizing the endless automatic polarization controlling device is depicted in Fig. S12a. To quantify the control system, the feedback signal, including the deviations caused by the voltage modulation, is recorded every 5.1 μs. A commercial polarization scrambler (General Photonics, PSY 201) is used to scramble the polarization state. Using the commercial polarization scrambler, the trajectories of the input polarization state can rotate around a given axis of the Poincaré sphere. In the main manuscript, we show the results of using our device to track the input polarization state which rotates counterclockwise. Here, we show the results when the input polarization state rotates around the clockwise (Fig. S12b). It indicates that our device still works in these cases.

Although the architecture (i.e. single-stage device) with one PSR and one MZI

which includes two EO phase shifters is sufficient for transforming any input SoP to the TE mode[2], the reset process is inevitable when the driving voltages reach the boundaries (see Fig. S13), resulting in momentary SoP mismatch and data interruption. While the reset does not happen in the multi-stage device which includes four EO phase shifters (see Fig. S14).

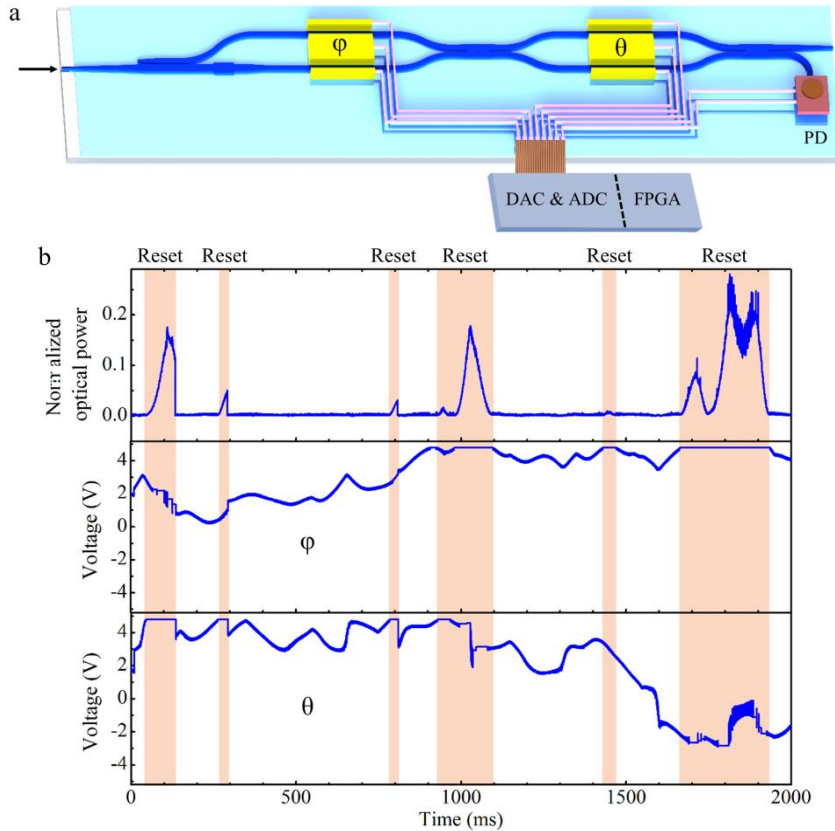

Fig. S13 **The experiment for characterizing single-stage polarization controlling device.** (a) The schematic of the single-stage polarization controlling device. (b) The normalized optical power of feedback signal, drive voltages of the $\theta$ and $\varphi$ phase shifters as the functions of time.

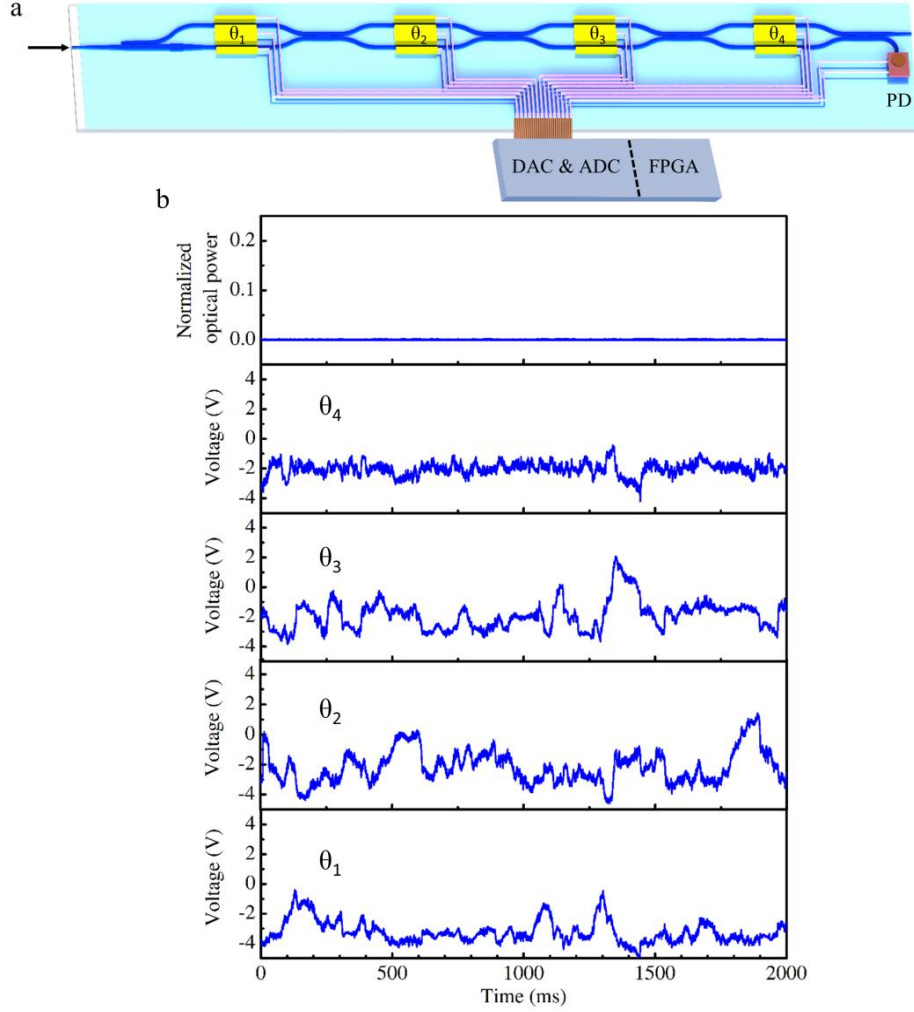

Fig. S14 **The experiment for characterizing multi-stage polarization controlling device.** (a) The schematic of the multi-stage polarization controlling device. (b) The normalized optical power of feedback signal, drive voltages of the corresponding four phase shifters as the functions of time.

## X. The summary of active bulky polarization management devices

Table 1 compares the performance of the active bulky polarization management devices. The active polarization management devices can be grouped into two kinds: the mechanical tuning device and the moving-part-free device. For the mechanical tuning device[3-5], generally, the insertion loss is low, but the tuning speed is low and its stability needs to be improved. The fiber squeezed with the piezo-electric actuator can achieve a relatively high tuning speed, but its drive voltage and power consumption are high[5].

The liquid crystal[6-9], lanthanum zirconate titanate waveplate[10], and magneto-optic crystal[11, 12] are often applied by the bulky moving-part-free device to control the polarization states. They can achieve a relatively high tuning speed and are more repeatable. While they still require high drive voltage or high power consumption. For example, the lanthanum zirconate titanate waveplate can achieve a response time of 1 μs, but the $V_\pi$ needs to be 200 V.

**Table 1** The comparison of the active bulky polarization management devices.

| Material | Tuning method | Drive Voltage ($V_\pi$) | Power for tuning | Speed (or response time) |
|---|---|---|---|---|
| Birefringent waveplate[3] | Mechanical rotation | - | - | 960 rad/s |
| Fiber | Squeezing with piezo-electric actuator[5] | 30V | ~20 W | 752 krad/s |
|  | twist[4] | - | - | Low |
| Liquid crystal[7] | Electro-optics | 40 V | - | ~100 μs |
| Lanthanum zirconate titanate waveplate[10] | Electro-optics | 200V | - | ~1 μs |
| Magneto-optic crystal[11, 13] | Magneto-optics | < 5 V | 1.3 W | ~150 μs |

## XI. Fabrication process

The proposed device was fabricated on X-cut LN-on-insulator (LNOI) wafer. The thicknesses of the LN and buried oxide layers are 360 nm and 4.7 μm, respectively. The fabrication process (Fig. S15) of optical component is detailed in the following:

(i)-(ii) Electron beam lithography (EBL) was first used to define the waveguide slab structures on the hydrogen silses quioxane (HSQ) resist.

(iii) The patterns were transferred to the top LN layer with an etching depth of 180 nm by inductively coupled plasma (ICP) dry etching.

(iv)-(v) The strip waveguides were defined on the LN layer with an etching depth of 360 nm using EBL and ICP dry etching. The strip waveguide is used for the edge coupler.

(vi) A $SiO_2$ layer with a thickness of 1 μm was deposited on the wafer as the upper-cladding by plasma-enhanced chemical vapor deposition (PECVD).

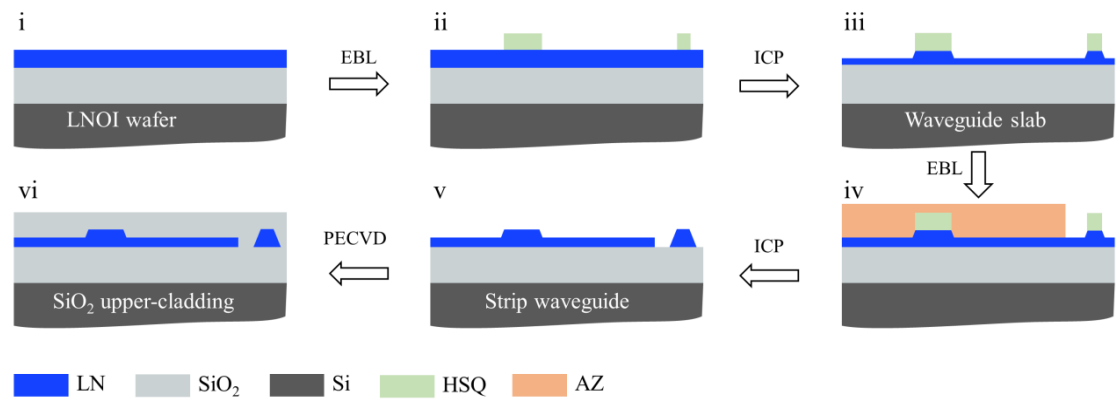

Fig. S15 The schematic of the fabrication process. EBL: electron beam lithography; EBE: electron beam evaporation; ICP: inductively coupled plasma etching; PECVD: plasma-enhanced chemical vapor deposition. E-beam resists include HSQ, AZ.